# Why 2D Semiconductors Generally Have Low Electron Mobility

Long Cheng, Chenmu Zhang, and Yuanyue Liu*

Texas Materials Institute and Department of Mechanical Engineering

The University of Texas at Austin, Austin, TX, 78712

*yuanyue.liu@austin.utexas.edu

**Abstract**: Atomically-thin (2D) semiconductors have shown great potential as the fundamental building blocks for next-generation electronics. However, all the 2D semiconductors that have been experimentally made so far have room-temperature electron mobility lower than that of bulk silicon, which is not understood. Here, by using first-principles calculations and re-formulating the transport equations to isolate and quantify contributions of different mobility-determining factors, we show that the universally low mobility of 2D semiconductors originates from the high "density of scatterings", which is intrinsic to the 2D material with parabolic electron band. The density of scatterings characterizes the density of phonons that can interact with the electrons, and can be fully determined from the electron and phonon band structures without knowledge of electron-phonon coupling strength. Our work reveals the underlying physics limiting the electron mobility of 2D semiconductors, and offers a descriptor to quickly assess the mobility.

**Main Text:** Single-unit-layer (2D) semiconductors have attracted great interest in recent years due to their unconventional properties and large potential in (opto)electronic applications.[1-5] The common 2D semiconductors include metal dichalcogenides ($MX_2$)[6], metal monochalcogenides (MX)[7,8], black phosphorus[9] and other group-15 materials[10]. Despite the diverse structure, chemical composition and properties of 2D semiconductors, they all have room-temperature electron mobility considerably lower than Si: the mobility (unless specially noted, the mobility in this paper always refers to the electron mobility at room temperature) of pure Si is ~ 1400 $cm^2V^{-1}s^{-1}$,[11] while those of 2D semiconductors that have been made experimentally are all < 400 $cm^2V^{-1}s^{-1}$.[4-6,9,12-16] This is an non-obvious fact; particularly if one considers the electron motion as random walk, then for the same mean square displacement ($R^2$), the diffusion constant $C_D$ (and thus the mobility) is higher in 2D than in 3D (because $C_D = R^2/nt$, where n = 4 for 2D while 6 for 3D, and $t$ is the diffusion time). Since the mobility is a key property for many applications, it is necessary to understand why it is universally low in 2D.

The mobility depends on the electronic structure and the scatterings. There are two common sources of scatterings for the 2D semiconductors in electronic applications: defects and phonons. The defects can come from the semiconductor itself or the environment. Although significant efforts have been made to minimize the defect-induced scattering, the improvement in mobility is limited; on the other hand, first-principles calculations have indicated that a low "intrinsic" mobility (i.e. the mobility in perfect material) for various 2D semiconductors[12,14,15,17-20], suggesting that the phonon-induced scattering may be the fundamental bottleneck. The low mobility has been attributed to different factors, depending on the material. For example, Cheng et al.[17] attributed the low intrinsic mobility of $MX_2$ to the strong scatterings by the longitudinal optical phonons (owing to the large Born charges); Fischetti and Vandenberghe[21] showed that the intrinsic mobility of the 2D semiconductors without the horizontal mirror symmetry is limited by the scatterings by the out-of-plane acoustic phonons. Giustino et al. attributed the lower mobility of 2D InSe than the 3D counterpart to the higher density of electronic states in 2D InSe.[15] Nevertheless, it remains unknown why 2D semiconductors generally have a lower electron mobility than Si.

In this work, we identify the high "density of scatterings" ($\overline{D^S}$), which is intrinsic to 2D semiconductors, as the origin for their universally lower electron mobility than that of Si. The paper is organized as follows: we first calculate the intrinsic mobility of various 2D semiconductors, using the state-of-the-art first-principles methods. Then we re-formulate the transport equations to isolate and quantify the contributions of relevant physical factors, which renders the $\overline{D^S}$ as the common source for low mobility of 2D semiconductors. Finally, we use model systems to show that the large $\overline{D^S}$ is inherent to the 2D material with parabolic electron band, and discuss possible strategies to improve the intrinsic mobility.

The mobility can be calculated as using Boltzmann transport equation:

$$\mu = |e| \frac{\sum_i \int_{BZ} \tau(i\mathbf{k}) v^2(i\mathbf{k}) |\frac{\partial f_{i\mathbf{k}}}{\partial E_{i\mathbf{k}}}| d\mathbf{k}}{\sum_i \int_{BZ} f_{i\mathbf{k}} d\mathbf{k}}, \qquad (1)$$

where $\tau(i\mathbf{k})$ is the relaxation time for the electronic state $i\mathbf{k}$ ($i$ is the band index, and $\mathbf{k}$ is the wavevector), $v$ is the velocity, $f$ is the occupation distribution at equilibrium, $E$ is the energy, and $e$ is the electron charge. In the following, we use $\mu$ to denote the intrinsic electron mobility at room temperature. The scattering rate ($1/\tau$) due to the phonons can be calculated as:

$$\frac{1}{\tau(i\mathbf{k})} = \frac{2\pi}{\hbar} \sum_{j,i'} \int_{BZ} \frac{d\mathbf{q}}{\Omega_{BZ}} |g(i\mathbf{k}, i'\mathbf{k}', j\mathbf{q})|^2 C_{i\mathbf{k},i'\mathbf{k}',j\mathbf{q}} \frac{1-f_{i'\mathbf{k}'}}{1-f_{i\mathbf{k}}} (1 - \frac{\mathbf{v}_{i\mathbf{k}} \cdot \mathbf{v}_{i'\mathbf{k}'}}{|\mathbf{v}_{i\mathbf{k}}| \cdot |\mathbf{v}_{i'\mathbf{k}'}|}), \qquad (2)$$

where

$$C_{i\mathbf{k},i'\mathbf{k}',j\mathbf{q}} = n_{j\mathbf{q}} \delta(E_{i\mathbf{k}} - E_{i'\mathbf{k}'} + \hbar\omega_{j\mathbf{q}}) \delta_{\mathbf{k}+\mathbf{q},\mathbf{k}'+\mathbf{G}} + (1+n_{j\mathbf{q}}) \delta(E_{i\mathbf{k}} - E_{i'\mathbf{k}'} - \hbar\omega_{j\mathbf{q}}) \delta_{\mathbf{k}-\mathbf{q},\mathbf{k}'+\mathbf{G}}. \qquad (3)$$

Here $i\mathbf{k}$, $i'\mathbf{k'}$, and $j\mathbf{q}$ denote the initial electronic state, the final electronic state, and the absorbed/emitted phonon, respectively. $\omega$ is the phonon frequency, $n$ is the Bose distribution, $\mathbf{G}$ is the reciprocal lattice vector, and $\Omega_{BZ}$ is the volume/area of the Brillouin zone. $C_{i\mathbf{k},i'\mathbf{k'},j\mathbf{q}}$ is the number of phonons that can scatter the electronic state $i\mathbf{k}$ to an infinitesimal energy window around $i'\mathbf{k'}$, divided by the range of energy window. Those phonons must satisfy the energy and momentum conservation of scattering, as required by the deltas in Eq. (3). As an example, Fig. 3a inset illustrates the phonon states that can scatter the conduction band minimum (CBM). $g(i\mathbf{k},i'\mathbf{k'},j\mathbf{q})$ is the electron-phonon coupling (EPC) strength, which can be calculated using the density functional perturbation theory (DFPT)[22,23] in conjunction with Wannier interpolation[24]. Note that here we treat the Fröhlich interaction using a 2D formula[19], different from that for 3D material[25]. The computation details can be found in the Supplemental Material[26] (SM).

Fig. 1 shows the calculated $\mu$ for various 2D semiconductors. The $\mu$ of bulk Si is also shown for comparison. Without strain: (1) among group-15 2D semiconductors, the black phosphorus (BP) has the highest $\mu$ (along armchair direction) of 303 cm$^2$V$^{-1}$s$^{-1}$, followed by the Bi (233 cm$^2$V$^{-1}$s$^{-1}$), while the others have the $\mu$ of only ~ 50 cm$^2$V$^{-1}$s$^{-1}$; (2) among 2D MX$_2$, WS$_2$ has the highest $\mu$ of 246 cm$^2$V$^{-1}$s$^{-1}$, followed by MoS$_2$ (183 cm$^2$V$^{-1}$s$^{-1}$), while the $\mu$ for all the others are below 100 cm$^2$V$^{-1}$s$^{-1}$; (3) for 2D MX, the $\mu$ are all below 40 cm$^2$V$^{-1}$s$^{-1}$. Our results are consistent with those in literature[14,16,17], and confirm that the $\mu$ of 2D semiconductors are well below that of Si. To explore the feasibility of achieving a higher $\mu$ by strain, we apply a 3% uniform tensile strain to each material and re-calculate the $\mu$. Indeed, the strain can significantly increase the $\mu$ for certain materials. For example, the $\mu$ of As increases from 49 cm$^2$V$^{-1}$s$^{-1}$ to 1267 cm$^2$V$^{-1}$s$^{-1}$, and that of Sb increases from 50 cm$^2$V$^{-1}$s$^{-1}$ to 1065 cm$^2$V$^{-1}$s$^{-1}$. Nevertheless, they are still lower than that of Si.

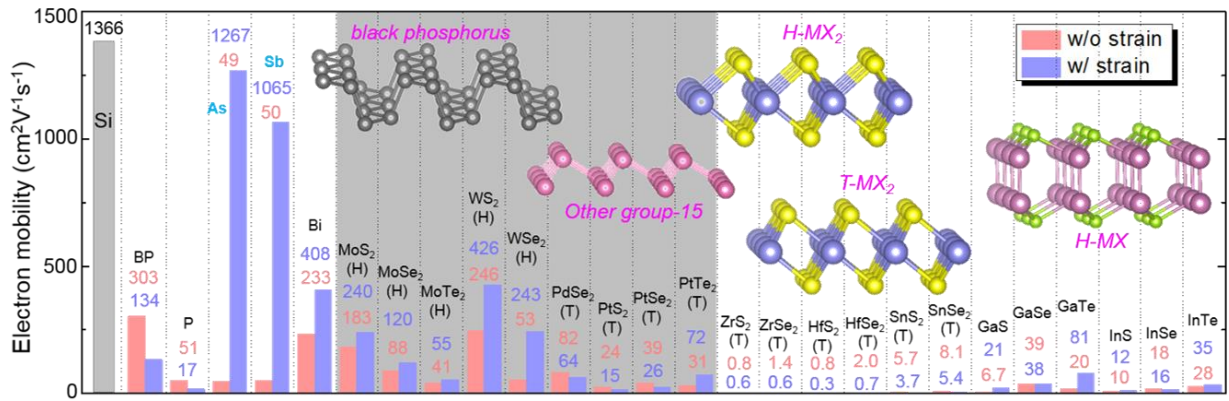

**Figure 1**. Intrinsic electron mobility at room temperature of the 2D semiconductors commonly studied in experiments. Results of materials without strain are shown in red, and those with a 3% uniform tensile strain are shown in blue. The inset figures show the side views of those 2D materials.

To understand the origin of the universally low $\mu$, we take a detailed look into Eq. (1) and (2). They indicate that the $\mu$ is determined by two types of factors: one is the EPC strength $g$, which represents the strength of electron-phonon interaction; and the other is "non-coupling" factors (e.g., electron/phonon energy, electron velocity, Fermi/Bose distribution), which are the properties of electrons/phonons alone. From the computation point of view, the EPC strength is usually difficult to calculate (requiring more time and CPUs), while the non-coupling factors can be computed relatively easily. To quantify and better understand the contributions of these factors, we seek to write the mobility as simple algebraic operations of physically-meaningful terms, so that their contributions can be easily isolated and quantified. To begin with, we first rewrite Eq. (2) as[16]:

$$\frac{1}{\tau(i\mathbf{k})} = \frac{2\pi}{\hbar} \left| g^*(i\mathbf{k}) \right|^2 D^S(i\mathbf{k}), \quad (4)$$

where

$$D^S(i\mathbf{k}) = \sum_{j,i'} \int_{BZ} \frac{d\mathbf{q}}{\Omega_{BZ}} C_{i\mathbf{k},i'\mathbf{k}',j\mathbf{q}} \frac{1-f_{i'\mathbf{k}'}}{1-f_{i\mathbf{k}}} (1 - \frac{\mathbf{v}_{i\mathbf{k}} \cdot \mathbf{v}_{i'\mathbf{k}'}}{|\mathbf{v}_{i\mathbf{k}}| \cdot |\mathbf{v}_{i'\mathbf{k}'}|}), \quad (5)$$

and

$$g^*(i\mathbf{k}) = \sqrt{\hbar / (2\pi D^S(i\mathbf{k}) \cdot \tau(i\mathbf{k}))}. \quad (6)$$

Eq. (4) can be interpreted as follows: the scattering rate for the electronic state $i\mathbf{k}$ is determined by the "effective" scattering/EPC strength for $i\mathbf{k}$ ($g^*(i\mathbf{k})$), and the "effective" density of scatterings available to $i\mathbf{k}$ ($D^S(i\mathbf{k})$). Indeed, the expression of $D^S(i\mathbf{k})$ shown by Eq. (5) confirms this interpretation: recall that the $C_{i\mathbf{k},i'\mathbf{k}',j\mathbf{q}}$ is the number of phonons that can scatter the electronic state $i\mathbf{k}$ (under the requirement of energy and momentum conservation) to an infinitesimal energy window around $i'\mathbf{k}'$, divided by the range of energy window; the $\frac{1-f_{i'\mathbf{k}'}}{1-f_{i\mathbf{k}}}$ accounts for the occupation of $i\mathbf{k}$ and $i'\mathbf{k}'$ (a less occupied $i'\mathbf{k}'$ makes the scattering more "effective"); and the $1 - \frac{\mathbf{v}_{i\mathbf{k}} \cdot \mathbf{v}_{i'\mathbf{k}'}}{|\mathbf{v}_{i\mathbf{k}}| \cdot |\mathbf{v}_{i'\mathbf{k}'}|}$ accounts for the change in the velocity direction (a large change makes the scattering more effective). The $D^S(i\mathbf{k})$ can be fully determined once the electron and phonon structures are given, without the knowledge of EPC. For the materials considered here, our calculations show that

$$D^S(i\mathbf{k}) \approx \sum_{j,i'} \int_{BZ} \frac{d\mathbf{q}}{\Omega_{BZ}} C_{i\mathbf{k},i'\mathbf{k}',j\mathbf{q}}, \quad (7)$$

because the carrier concentration is low (Fermi level is in the middle of the band gap, as required by the definition of intrinsic mobility), and the band structures are quasi-isotropic. Consider an extreme situation where there is only one phonon band with constant energy: in this case, the $i\mathbf{k}$ state can be scattered to any electronic state with the energy of $E_{i\mathbf{k}} +/- \hbar\omega$, and the $D^S(i\mathbf{k})$ will approximately be the sum of density of electronic states at these two energies multiplied by the number of phonons with energy $\hbar\omega$. In general, a higher density of electronic states and a larger number of phonons (which can be realized by lowering phonon energy) increase the chance of scattering, resulting in a higher $D^S$, and vice versa.

Substituting Eq. (4) to Eq. (1), we get:

$$\mu = \frac{\hbar|e|}{2\pi} \frac{\sum_i \int_{BZ} \left[ |g^*(i\mathbf{k})|^2 D^S(i\mathbf{k}) \right]^{-1} v^2(i\mathbf{k}) | \frac{\partial f_{i\mathbf{k}}}{\partial E_{i\mathbf{k}}} | d\mathbf{k}}{\sum_i \int_{BZ} f_{i\mathbf{k}} d\mathbf{k}}. \quad (8)$$

With further derivations shown in the SM, we can express the mobility as simple algebraic operations of physically-meaningful terms:

$$\mu = \frac{|e|\hbar}{2\pi} \cdot \frac{1}{\left(\overline{m^*} \cdot \overline{D^S}\right)} \cdot \frac{1}{\left(\overline{g^*}\right)^2}, \quad (9)$$

where $\overline{m^*}$ is the average effective mass[16], $\overline{D^S}$ is the average effective density of scatterings (which collects the $D^S$ of all the electronic states), and $\overline{g^*}$ is the average effective EPC strength (which collects the EPC strength of all the scatterings). They are defined as:

$$\overline{m^*} = \frac{\sum_i \int f_0 d\mathbf{k}}{\sum_i \int v^2(i\mathbf{k}) | \frac{\partial f_0}{\partial E_{i\mathbf{k}}} | d\mathbf{k}}, \quad \overline{D^S} = \frac{\sum_i \int_{BZ} v^2(i\mathbf{k}) | \frac{\partial f_0}{\partial E_{i\mathbf{k}}} | d\mathbf{k}}{\sum_i \int_{BZ} \frac{1}{D^S(i\mathbf{k})} v^2(i\mathbf{k}) | \frac{\partial f_0}{\partial E_{i\mathbf{k}}} | d\mathbf{k}}, \quad \text{and } \overline{g^*} = \sqrt{\frac{|e|\hbar}{2\pi} \cdot \frac{1}{\overline{m^*} \cdot \overline{D^S} \cdot \mu}}. \quad (10)$$

Note that the $\overline{D^S}$ can be fully determined once the electronic and phonon structures are given.

The reformulation of Eq. (1) and (2) to Eq. (9) enables us to isolate, and more importantly, quantify the contributions of physical factors determining the $\mu$. Dividing Eq. (9) by the corresponding quantities of Si then taking the logarithm, we get:

$$\log(\mu/\mu_{Si}) = \log\left(\overline{m^*_{Si}}/\overline{m^*}\right) + \log\left(\overline{D^S_{Si}}/\overline{D^S}\right) + \log\left[\left(\overline{g^*_{Si}}\right)^2 / \left(\overline{g^*}\right)^2\right]. \quad (11)$$

Therefore, the $\mu$ difference between Si and any material (characterized by $\log(\mu/\mu_{Si})$) can be fully decomposed to the differences in $\overline{m^*}$, $\overline{D^S}$, and $\overline{g^*}$, which are shown in Figure **2**. We find that: (1) the $\overline{D^S}$ of all the 2D materials studied here (both unstrained and strained) is higher than that of Si, while the $\overline{g^*}$ and $\overline{m^*}$ can be larger or smaller depending on the material. Blue phosphorus (P) has the lowest $\overline{D^S}$ in these 2D systems, which is however still 14 times higher than that of the Si. This suggests that the reason why these 2D materials have universally lower $\mu$ than Si is that they all have higher $\overline{D^S}$; (2) the $\overline{g^*}$ and $\overline{m^*}$ can further enhance/suppress the $\mu$ difference. For example, many T-MX$_2$ and all MX have a larger $\overline{g^*}$ than Si (due to their strong electrical polarization oscillation caused by longitudinal optical phonons[17]), which further decreases their $\mu$. In contrast, the H-MX$_2$ and group-15 semiconductors (except blue phosphorus, P) have a smaller $\overline{g^*}$ than Si, which is however not enough to compensate the effect of larger $\overline{D^S}$; therefore the $\mu$ of these materials is still lower than that of Si.

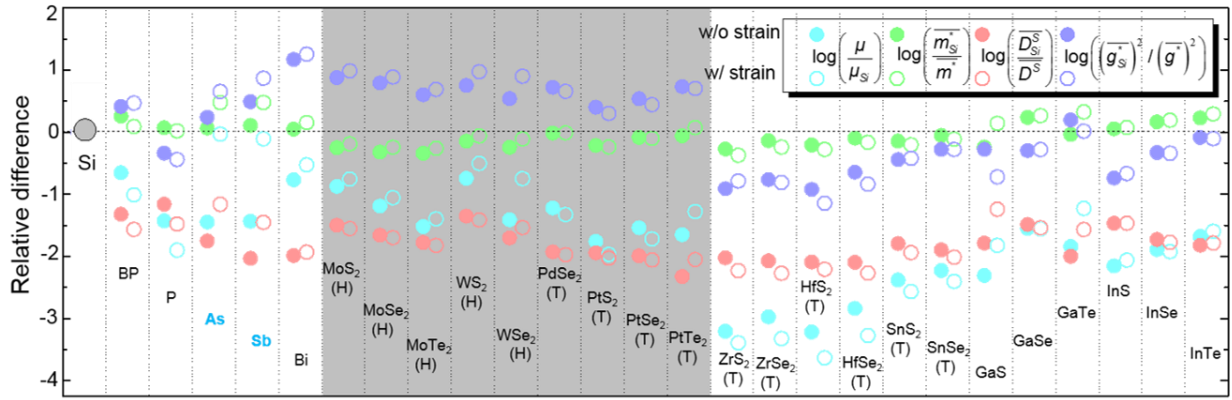

**Figure 2**. Differences in mobility ($\mu$), average effective mass ($\overline{m^*}$), average effective density of scatterings ($\overline{D^S}$), average effective electron-phonon coupling strength ($\overline{g^*}$), between 2D semiconductors and Si, expressed in the form based on Eq. (11). Results of materials without strain are shown by solid circles, and those with a 3% uniform tensile strain are shown by hollow circles.

To explore whether the higher $\overline{D^S}$ is limited to the 2D materials studied here, we construct a model 2D system that has only one parabolic electron band and three identical linear phonon bands with tunable curvature and slope. The phonons have zero energy when the wavelength becomes infinitely large, i.e. they are acoustic. If we take the curvature (or effective mass, $m^*$) of the electron band from the Si CBM, and take the slope of the phonon band (or sound velocity, $v_s$) from the Si longitudinal acoustic phonon band near the Γ point, and set the Fermi level to be at the same position (relative to the CBM) as that in perfect

Si, then we find that the $\overline{D^S}$ of the model 2D system is much larger than that of Si (in fact, as shown in Fig. 3a, the $D^S(i\mathbf{k})$ is also significantly larger over a wide range of energy). Fig. 3b shows how the $\overline{D^S}$ of 2D system changes with the parameters $m^*$ and $v_s$. A smaller $m^*$ and a larger $v_s$ give a lower $\overline{D^S}$, consistent with the expectation that the lower density of electronic states and a higher phonon energy result in a smaller $\overline{D^S}$. However, in order to get a $\overline{D^S}$ as low as that of Si, $m^*$ and $v_s$ have to take extreme values. For example, if the $v_s$ takes the highest value of known materials (21.6 km/s in graphene[30]), then $m^*$ must be < 0.1 $m_e$, a very small value for semiconductor. Note that this model considers only three acoustic phonon bands, while in reality there often exist more phonon bands that can give more scatterings. Therefore, the $\overline{D^S}$ value derived from this model should be interpreted as the lower limit. Since the lower limit is already impractical to be comparable with the Si $\overline{D^S}$, we hence conclude that 2D semiconductor practically always has a larger $\overline{D^S}$ than Si, which in turn limits its mobility.

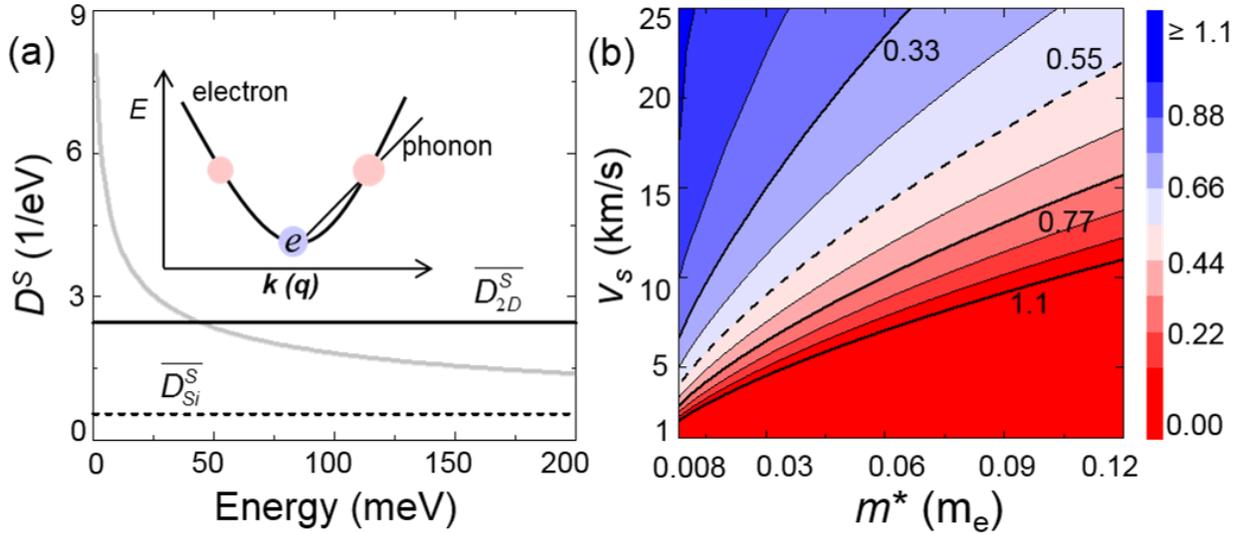

Figure 3. (a) Gray line: effective density of scatterings $D^S$ as a function of electronic energy (in reference to the energy of conduction band minimum) for a model 2D system (see the main text), with the effective mass ($m^*$) and sound velocity ($v_s$) taken from Si. The average effective density of scatterings, $\overline{D^S}$, is shown by the solid black line for the model 2D system and by the dotted line for Si. The inset illustrates the phonon state (red) that can scatter the electronic state in blue. (b) The $\overline{D^S}$ of the model 2D system as a function of $m^*$ and $v_s$. The Si value is marked for comparison.

The universally higher $\overline{D^S}$ of 2D system can be attributed to its higher density of electronic states and more phonons at low energies. For a parabolic electron band, the density of states scales as $\sqrt{E-E_c}$ (where the $E_C$ is the energy of CBM) in 3D, while is an energy-independent constant in 2D; thus the density of electronic states near CBM is always larger in 2D than in 3D (see Fig. S3). Similarly, for a linear phonon band, the density of states scales as $(\hbar\omega)^2$ in 3D, while as $(\hbar\omega)$ in 2D; thus the 2D system always has more phonons at low energy than 3D system (see Fig. S3). In addition to the dimensionality effect, the density of phonons is also affected by material-specific properties such as the atomic masses and the bonding character (a larger mass and a weaker bonding would give a higher density of phonons). Similarly, the density of electronic states is also affected by the material-specific properties such as delocalization degree of the atomic orbitals forming the conduction band. Although it is impractical to make the $\overline{D^S}$ in 2D to be smaller than that in Si, it is still useful to explore how to decrease the 2D $\overline{D^S}$ to improve the mobility. This can be achieved by decreasing the density of electronic states and increasing the phonon energy. The effect of density of electronic states can be seen by comparing the unstrained and strained As: as shown in Fig. S4a, the tensile strain shifts up some of the valleys, decreasing the number of the valleys near the conduction band edge from 6 to 1. Consequently, the density of electronic states decreases (Fig. S4b) and the $\overline{D^S}$ decreases (Fig. 2). This is one of the reasons why the As $\mu$ significantly increases by 25 times (from 49 $cm^2V^{-1}s^{-1}$ to 1267 $cm^2V^{-1}s^{-1}$) with a 3% tensile strain. Similar effect exists for Sb (which increases by 20 times from 50 $cm^2V^{-1}s^{-1}$ to 1065 $cm^2V^{-1}s^{-1}$), and has been reported for other systems[16,18,31]. The effect of phonon energy can be seen by comparing the unstrained and strained $MoS_2$: as shown in Fig. S4d, the tensile strain sharpens the electron band, thereby decreasing the density of electronic states (Fig. S4e). At first thought this should decrease the $\overline{D^S}$, in apparent contradiction with the observed increase. Examining the change in phonon structure finds that the increase of $\overline{D^S}$ originates from the decrease of phonon energy, due to the phonon softening by tensile strain (Fig. S4f).

Finally, we point out the $\overline{D^S}$ is not the only factor that determines the $\mu$. According to Eq. (9), a small $\overline{g^*}$ can yield a high mobility, and as shown in Fig. 2, the $\overline{g^*}$ of some 2D semiconductors can be lower than that of Si. Therefore, it is possible to find a 2D semiconductor with a small enough $\overline{g^*}$ that overcomes the limitation of $\overline{D^S}$, which thus has a higher mobility than Si. The $\overline{g^*}$ can be further tuned by strain (Fig. 2) or substrate[19]. We also note that the cost of calculating $\mu$ can be reduced with a good efficiency-accuracy tradeoff using electron-phonon averaged (EPA) approximation approach as proposed in Ref.[32]. The main idea of the EPA approximation is to turn the complex momentum-space integration in Eq. 2 into an integration over energies. This is accomplished by replacing momentum-dependent quantities in Eq. 2 by their energy-dependent averages. The EPA has been demonstrated to be successful in computationally screening the half-Heusler family of compounds for thermoelectric power generation applications[32]. Lastly, we note that the high density of low-energy phonons in 2D system may have implications for thermal and thermoelectric properties and applications; and the concept of density of scatterings may also apply to phonons and could also be an important factor in determining the difference in the thermal transport across different materials.

In summary, we show that the universally low electron mobility of 2D semiconductors at room temperature originates from their high density of scatterings, which is inherent to the 2D material with parabolic electron band. The density of scatterings characterizes the density of phonons that can interact with the electrons, and can be easily calculated from the electron and phonon band structures without the knowledge of electron-phonon coupling strength, thereby providing a physically intuitive and computationally easy descriptor to explain the mobility or search for high-mobility semiconductors.

**Acknowledgments**

This work is supported by Welch Foundation (Grant No. F-1959-20180324) and the startup grant from UT Austin. This work used computational resources located at the National Renewable Energy Laboratory sponsored by the DOE's Office of Energy Efficiency and Renewable Energy, and the Texas Advanced Computing Center (TACC) at UT Austin.